\documentclass[twocolumn,preprintnumbers,amsmath,amssymb]{revtex4}

\usepackage{graphicx}
\usepackage{dcolumn}
\usepackage{bm}
\usepackage{color} 

\begin{document}


\title{  Experimental evidence of $T_c$ enhancement without the influence of spin fluctuations: NMR study on LaFeAsO$_{1-x}$H$_x$ under a
pressure of 3.0 GPa
 }

\author{ N. Kawaguchi$^{1}$, N. Fujiwara$^1$\footnote {Email:naoki@fujiwara.h.kyoto-u.ac.jp},  S. Iimura$^{2, 3}$, S. Matsuishi$^{2, 3}$,
and H. Hosono$^{2, 3}$ }

\affiliation{$^1$Graduate School of Human and Environmental Studies, Kyoto University, Yoshida-Nihonmatsu-cyo, Sakyo-ku, Kyoto 606-8501,
Japan}

\affiliation { $^2$Material and Structures Laboratory (MSL), Tokyo Institute of Technology, 4259 Nagatsuda, Midori-ku, Yokohama 226-8503,
Japan \ \\ $^3$Frontier Research Center (FRC), Tokyo Institute of Technology, 4259 Nagatsuda, Midori-ku, Yokohama 226-8503, Japan }





\begin{abstract}

The electron-doped high-transition-temperature ($T_c$) iron-based pnictide superconductor LaFeAsO$_{1-x}$H$_x$ has a unique phase
diagram: Superconducting (SC) double domes are sandwiched by antiferromagnetic phases at ambient pressure and they turn into a single dome
with a maximum $T_c$ that exceeds 45K at a pressure of 3.0 GPa. We studied whether spin fluctuations are involved in increasing $T_c$
under a pressure of 3.0 GPa by using the $^{75}$As nuclear magnetic resonance (NMR) technique. The $^{75}$As-NMR results for the powder samples show that
$T_c$ increases up to 48 K without the influence of spin fluctuations. This fact indicates that spin fluctuations are not involved in raising $T_c$,  which implies that other factors, such as orbital degrees of freedom, may be important for achieving a high $T_c$ of almost 50 K.

\end{abstract}

\maketitle

The phase diagram of the electron-doped high-transition-temperature ($T_c$) iron-based pnictide LaFeAsO$_{1-x}$H$_{x}$ (H-doped La1111
series) is unique owing to the capability of electron doping: (i) It exhibits a superconducting (SC) phase with double domes covering a
wide H-doping range from $x=0.05$ to $x=0.44$ $^{1}$, (ii) the SC phase is sandwiched by antiferromagnetic (AF) phases appearing in heavily
and poorly electron-doped regimes [see Fig. 1(a)] $^{2}$, and (iii) the application of pressure transforms the double domes into a single dome
$^{1, 3}$. Intriguingly, upon applying pressure, the minimum $T_c$ at ambient pressure becomes the maximum $T_c$ of over 45 K $^{1}$, as shown by the solid arrow in Figs. 1(a) and 4 as described in detail below.

\begin{figure} \includegraphics{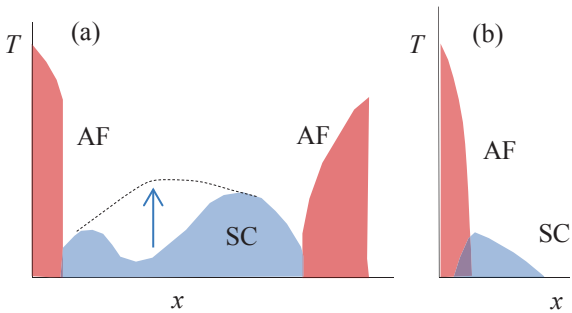} \caption{\label{fig:epsart} (color online) Schematic phase diagrams of (a) H-doped La1111
series LaFeAsO$_{1-x}$H$_x$, and of (b) Ba122 series such as Ba(Fe$_{1-x}$Co$_x$)$_2$As$_2$.  AF and SC represent antiferromagnetic and
superconducting phases, respectively. The arrow in (a) shows the enhancement of $T_c$ that occurs between
ambient pressure and 3.0 GPa. } \end{figure}

The unique features established in this compound have not been observed so far in other iron-based pnictides, such as the Ba122 and Na111
series, which have been investigated intensively from an early stage because they are available as large crystals. The electronic phase
diagram of the Ba122 series is similar to that of high-$T_c$ cuprates $^{4}$. The analogy is reminiscent of the importance of AF
fluctuations in iron-based pnictides. The spin-fluctuation-mediated mechanism is a major
candidate for the high-$T_c$ mechanism. In fact, the SC phase of the Ba122 series partially overlaps the AF phase, in other words, the SC and
AF states are compatible, and the maximum $T_c$ occurs close to the phase boundary [see Fig. 1(b)] $^{5, 6}$. Because of this special
location, $T_c$ is enhanced and low-energy AF fluctuations simultaneously become predominant as the doping level approaches the AF phase $^{7, 8}$. The
Na111 series has a phase diagram similar to that of the Ba122 series; however, the SC phase overlaps the AF phase over a wide doping
range and even extends to the undoped material $^{9}$. By tuning pressure, $T_c$ and AF fluctuations are
found to be related in a similar manner as in the Ba122 series $^{10}$. A pressure-enhanced $T_c$ occurs in the 11 series FeSe, which is
superconducting and has no magnetic orders at ambient pressure. At first sight, the series seems to be free from the antiferromagnetism;
however, at pressures exceeding 1 GPa, the SC phase is adjacent to the AF or AF+SC phase $^{11}$. In fact, the influence of AF fluctuations
is observable in the SC phase even at ambient pressure $^{12}$.

\begin{figure*} \includegraphics{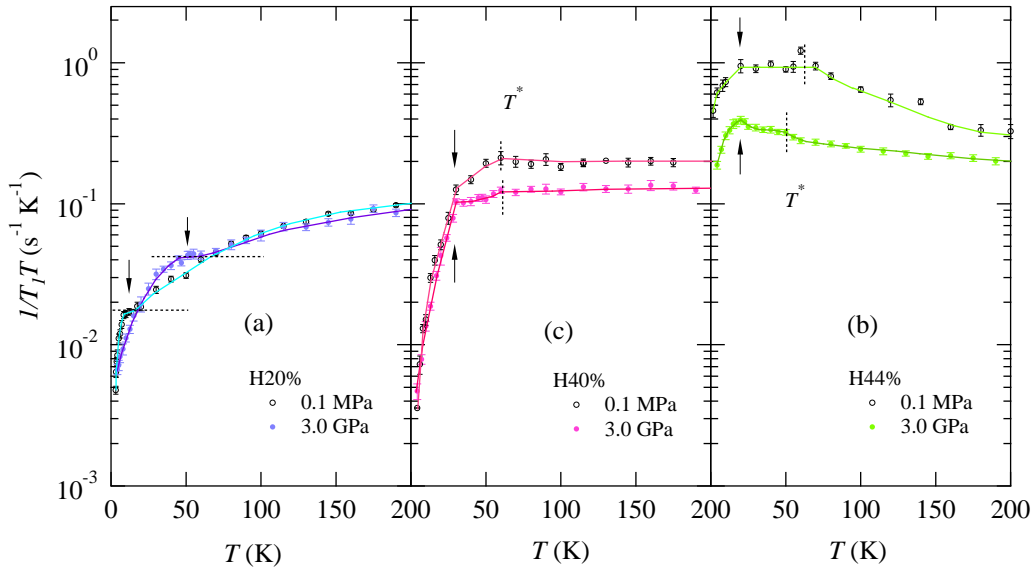} \caption{\label{fig:epsart} (color online) Nuclear magnetic relaxation rate divided by
temperature $1/T_1T$ for $^{75}$As for (a) $x=0.20$, (b) $x=0.40$, and (c) $x=0.44$. Arrows represent $T_c$  and solid curves are guides to the eye. Horizontal dotted lines in (a) represent plateaus just above $T_c$. For $x=$0.40 and 0.44, an anomaly of $1/T_1T$ appears at $T^*$ in a paramagnetic phase (see vertical dotted lines).  } \end{figure*}

\begin{figure*} \includegraphics{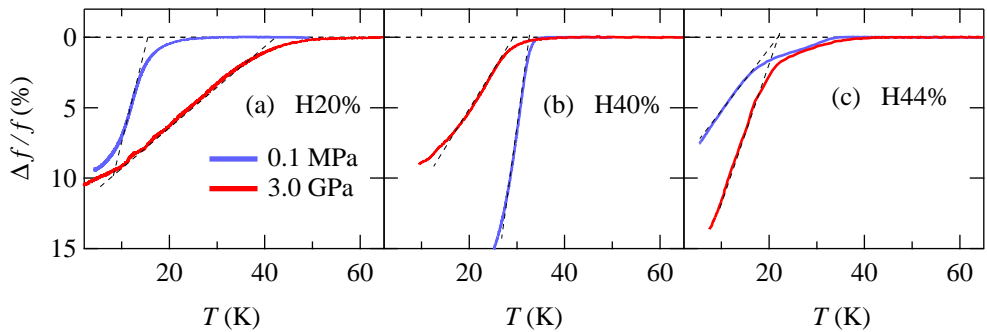} \caption{\label{fig:epsart} (color online) Detuning of resonance frequency for
(a) $x=0.20$, (b) $x=0.40$, and (c) $x=0.44$.  $T_c$'s determined from $1/T_1T$ [see arrows in Figs. 1(a)-1(c)] are in good agreement with the extrapolations of dashed lines. } \end{figure*}

AF fluctuations seem to play a key role in raising $T_c$ for various iron-based pnictides; however, the scenario does not work well for
LaFeAsO$_{1-x}$F$_{x}$, because for $x=0.14$,  $T_c$ increases up to 40 K at 3.0 GPa with no predominant AF fluctuations $^{13, 14}$. The
La1111 series under high pressure is the only material available for investigating the magnetic properties of pnictides with $T_c$ in the
range of 45-50 K. In fact, the Sm1111 series marks the highest $T_c$ ($T_c$=55 K) in all types of iron-based pnictides $^{15}$; however, it
includes magnetic Sm ions, which hinders the investigation of the magnetic properties of iron-basal planes. The rise of $T_c$ without AF
fluctuations was observed only for $x=0.14$ (see the dashed arrow in Fig. 4), because AF fluctuations remain in a lower doping range than
$x=0.14$ and unfortunately $x=0.14$ is nearly the highest level of F doping. So far as nuclear-magnetic-resonance (NMR) studies on
LaFeAsO$_{1-x}$F$_x$ are concerned, the maximum doping level is less than $x=0.14-0.15$ $^{16, 17}$.  To establish the breakdown of this
scenario over a wide doping range (0.20 $\leq x \leq$ 0.44) that covers the second SC dome, we applied $^{75}$As ($I$=3/2) NMR to the powder samples of the H-doped La1111 series at 3.0 GPa.

\begin{figure} \includegraphics{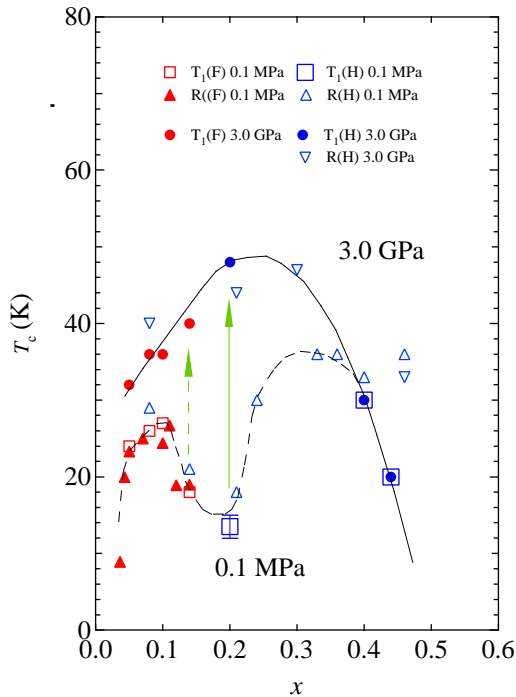} \caption{\label{fig:wide} (color online)  Phase diagram of LaFeAsO$_{1-x}$H$_{x}$ (0.05 $\leq x
\leq$ 0.5) and LaFeAsO$_{1-x}$F$_{x}$ (0.05 $\leq x \leq$ 0.14).  Solid and open triangles represent $T_c$ determined by the
resistivity at ambient pressure, and open squares represent $T_c$ determined by the relaxation time ($T_1$). Downward pointing triangles and
solid circles represent $T_c$ determined by the resistivity and $T_1$ at 3.0 GPa, respectively. The dashed and solid arrows indicate
the enhancement of $T_c$ that occurs when pressure is applied.  } \end{figure}

 We applied a pressure of 3.0 GPa to samples with $x=$0.20, 0.40, and 0.44. The pressure was applied by using NiCrAl-CuBe hybrid
 piston-cylinder-type cells. We used a mixture of F-70 and F-77 fluorinate as the pressure mediation liquid. The details of the pressure
 cells are given in Ref. 18. A coil wound around the samples inside the pressure cell and capacitors equipped with a NMR probe form
 the tank circuit, which serves to detect the detuning of the resonance frequency, namely the ac susceptibility and to detect the NMR
 signal as well. The NMR measurements were performed using a conventional coherent pulsed-NMR spectrometer.
The $^{75}$As-NMR spectra show a broad powder pattern with double edges$^{14}$, which originates from the second-order quadrupole effect under a magnetic field. The relaxation time ($T_1$) for $^{75}$As was measured by using the saturation-recovery method at the lower-field edge in the field-swept NMR spectra. The low-field edge appears at about 48.2 kOe for an NMR frequency of 35.1 MHz. The signals at the low-field edge come from the powder samples with the iron-basal planes parallel to the applied field. Figures 2(a)-2(c) show the evolution of the relaxation rate ($1/T_1$) divided by temperature ($T$),  ( i.e., $1/T_1T$). Here, we chose $T_c$ as the onset of $1/T_1T$ as plotted as arrows in Figs. 2(a)-2(c). For $x=$0.20, $1/T_1T$ just above $T_c$ exhibited plateaus as shown by dotted lines in Fig. 2(a), and $T_c$ was remarkably enhanced upon applying pressure.  For $x=$0.40 and 0.44, an anomaly was observed in a paramagnetic state as marked as $T^*$, and both $T^*$ and $T_c$ were unchanged upon applying pressure. For $x$=0.44, Curie-Weiss-like behavior, which implies AF fluctuations, appears above $T^*$ at 0.1 MPa; however, this behavior has no appreciable effect on $T_c$.

Note that $T_c$ was determined under the applied field. In general, $T_c$ decreases more or less under an applied field; however, the decrease is significantly suppressed because we measured $1/T_1T$ for the powder samples with the iron-basal planes parallel to the applied field.  In fact, the values of $T_c$ are in good agreement with those determined from the detuning of resonance frequency at zero field. In this measurement, $T_c$ can be determined from the extrapolation as shown as dashed lines in Figs. 3(a)-3(c). Figure 4 shows the doping dependence of $T_c$ determined from the resistivity $^{1}$ and $1/T_1T$. The data at 3.0 GPa for the low-doping regime are cited from the results of the F-doped La1111 series $^{14, 19}$. As highlighted by the solid arrow in Fig. 4, $T_c$ increases to 48K at 3.0 GPa, which is comparable to the highest $T_c$ ($\sim$55 K) for all types of iron-based superconductors marked in the Sm1111 series.

In general, $1/T_1T$ of d-electrons systems is determined by spin correlations and is expressed by using the imaginary part of the dynamical spin susceptibility $Im\chi(q, \omega)$ as $1/T_1T \propto Im\chi(q, \omega_N)/\omega_N$, where $\omega_N$ is the angular frequency of nuclei. When the interaction between electrons is significantly strong, namely spin fluctuations are predominant, Curie-Weiss-like behavior is derived, whereas, when the interaction is week, $1/T_1T$ is determined by the density of states (DOS) at Fermi surfaces.  Unlike other pnictides, Curie-Weiss-like behavior is not observable for $x=0.20$ as seen in Fig. 2(a). Another example is K$_y$Fe$_{2-x}$Se$_2$ with $T_c=30$ K $^{20}$: The compound  exhibits similar $T$ dependence to the La1111 series. The results in Fig. 2(a) demonstrate that $T_c$ is enhanced without appreciable low-frequency AF fluctuations, which is the most important result for this study. The absence of AF fluctuations has also been confirmed from the neutron-scattering measurements in both LaFeAsO$_{1-x}$H$_x$ $^{21}$ and LaFeAsO$_{1-x}$F$_x$ $^{22}$: An inelastic-scattering peak is absent for $x=$0.20, despite the fact that it is unambiguously observable near the AF phase.

Herein, the $T$ dependence of $1/T_1T$ is attributable to the DOS. The monotonous $T$ dependence at high temperatures is attributed to the DOS involved only at high temperatures. In fact, the photoemission spectroscopy measurements demonstrate that the DOS for LaFeAsO$_{1-x}$F$_x$ decreases with decreasing temperature $^{23}$. This scenario is approved by quantitative evaluation of $1/T_1T$ just at the plateau. We evaluate $1/T_1T$ using the Korringa relation $^{24, 25}$ for d-electron alloys $^{26}$:

 \begin {equation}
 1/T_1T = \frac{\pi}{\hbar}(2\hbar\gamma_{N}A_{hf})^2\sum_{i}n_{i}(\varepsilon_F)^2 \frac{k_{B}}{(1-\alpha_Q)^2},
 \end {equation} where $\gamma_{N}$, $A_{hf}$ and $n_i(\varepsilon_F)$ represent the gyromagnetic ratio of $^{75}$As (7.292 MHz/10 kOe),
 the hyperfine coupling constant and DOS at Fermi surfaces for $i=d_{xy}, d_{yz}$, and $d_{zx}$ orbits, respectively. The
 factor $\alpha_Q$ is $I\chi(Q)$ where $I$ is the interaction between electrons and $\chi(Q)$ is the susceptibility without the
 interaction at dominant wave number $Q$. The value of $A_{hf}$ has been estimated to be $\sim$25 kOe/$\mu_B$ from the $K-\chi$ plot
 $^{27}$. The theoretically calculated values of  $n_i(\varepsilon_F)$ for $x=$0.20 at ambient pressure are 0.62, 0.92, and 0.92 (eV$^{-1}$),
 respectively, for $i=d_{xy}, d_{yz}$ and $d_{zx}$ orbits $^{1}$. These values result in $1/T_1T = 4.96\times10^{-4}
 \frac{1}{(1-\alpha_Q)^2}$ (s$^{-1}$K$^{-1}$). The plateau of $1/T_1T$ at 0.1 MPa indicates 0.018 (s$^{-1}$K$^{-1}$), and thus
 $\alpha_Q=0.83$. The value is in good agreement with the theoretically estimated value $\alpha_Q=0.94$ for $x=0.20$ $^{28}$.  At high temperatures, $1/T_1T$ at 3.0 GPa is almost the same as that at 0.1 MPa, which suggests
 that $\alpha_Q$ is insensitive to pressure and thus $\alpha_Q$ is not a key parameter for increasing $T_c$.

\begin{figure} \includegraphics{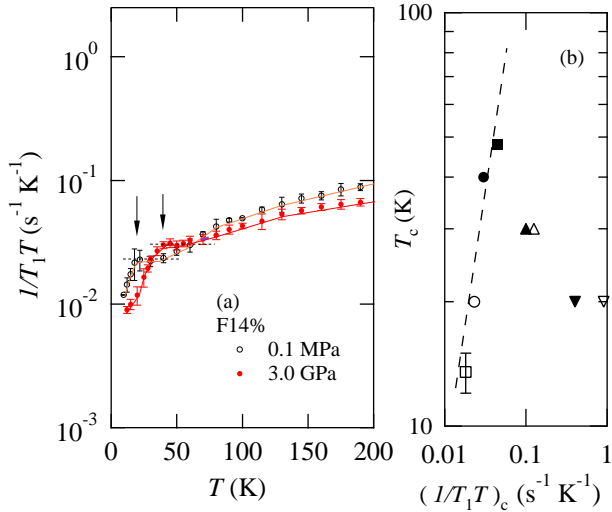} \caption{\label{fig:epsart} (color online)  (a) $1/T _1T$ for LaFeAsO$_{1-x}$F$_x$ ($x=$0.14).
Arrows represent $T_c$.   (b)  $T_c$ vs $1/T_1T$ at $T_c$ for LaFeAsO$_{1-x}$F$_x$ and LaFeAsO$_{1-x}$H$_x$.
Circles represent the data for $x=$0.14 (F doping), and squares, regular triangles, and downward pointing triangles represent those for $x=$0.20, 0.40, and 0.44 (H doping), respectively. Open and solid marks represent the measurements at 0.1 MPa and 3.0 GPa, respectively. The dashed line is a guide to the eye.} \end{figure}

Figure 5(a) shows the plateaus observed for 14\% F-doped samples $^{13, 14}$. The $T_c$ enhancement upon applying pressure is highlighted by the
dashed arrow in Fig. 4. At 0.1 MPa,  the values of $1/T_1T$ at $T_c$, $(1/T_1T)_c$ are 0.023 and 0.018 (s$^{-1}$K$^{-1}$) for 14\% F- and 20\% H-doped samples, respectively. At 3.0 GPa, these numbers increase to 0.030 and 0.044, respectively. These data are plotted in
Figure 5(b).  The data for 40\%  and 44\% H-doped samples are also plotted for comparison. As seen from the figure, $T_c$ correlates with $(1/T_1T)_c$ only for 14\% F- and 20\% H-doped samples reflecting $n_i(\varepsilon_F)$ in Eq. (1).

 Our results for the La1111 series demonstrate that $T_c$ is not directly affected by AF fluctuations as clearly seen from Figs. 2(a) and 2(c). Note that the opposite conclusion
 was derived for the Na111 series, such as NaFe$_{1-x}$Co$_{x}$As $^{10}$. In the Na111 series, $T_c$ follows AF fluctuations when pressure
 is applied. At first sight, the results of the Na111 series seem to contradict the results reported herein. One may classify the La1111 series as an exotic and exceptional compound among iron-based
 pnictides. However, all facts including the La1111 and Na111 series are consistent if the superconductivity is not directly affected by
 antiferromagnetism or AF fluctuations. For all iron-based pnictides without exception, $1/T_1T$ becomes stronger as the doping level approaches the AF-phase boundary, but the $T_c$ optimal doping level is not always located on the AF-phase boundary and depends on the compounds, which causes an apparent discrepancy.

 Roughly speaking, $T_c$ is proportional to the DOS and the paring interaction. The enhancement of the paring interaction is hardly expected for the AF-fluctuation-mediated scenario. As another candidate, the orbital-fluctuation-mediated scenario $^{28}$ would be promising. In this case, orbital fluctuations are difficult to observe in $1/T_1T$ at a doping level where the structural or nematic phase is absent, and thus the increase in $T_c$ is observable via the DOS alone in $1/T_1T$. To investigate whether the paring interaction is enhanced simultaneously as well as the DOS, further theoretical investigations are needed; however, the results of Fig. 5(b) would give an important clue.

 One may consider another scenario where the pairing interaction originates from AF fluctuations, but AF fluctuations are suppressed at ambient pressure by some competing interactions such as orbital and/or charge interactions. On the basis of this scenario,  $T_c$ could be suppressed and the double-dome structure could be observed as observed in some high-$T_c$ cuprates. Pressure could nullify the competition, and $T_c$ may increase under pressure even if AF fluctuations are not enhanced. However, the competing orders, which could cause appreciable suppression of $T_c$,  have not been observed so far in a wide range around $x=0.14-0.20$. Furthermore, AF fluctuations tend to decrease by applying pressure as observed in the poorly F-doped regime $^{29}$  or sufficiently H-doped regime [See Fig. 1(c)]. At present, there is no experimental evidence to support this scenario.

 In conclusion, we have observed in LaFeAsO$_{1-x}$H$_x$ that $T_c$ for $x=0.20$ marks a high $T_c$ of 48 K upon applying pressure without
 the influence of AF fluctuations. For $x=0.44$ (near the second AF-phase boundary), $T_c$ remains unchanged  without depending on the
 magnitude of AF fluctuations. These results suggest that the superconductivity has no direct connection with AF fluctuations. So far as the results of $1/T_1T$ are concerned, the increase in $T_c$ up to 48 K originates from an enhancement of the DOS just above $T_c$.

The authors thank H. Kontani and Y. Yamakawa for discussion.







\end{document}